\documentclass[prl, amsfonts, twocolumn, nofootinbib, showpacs]{revtex4}
\usepackage{graphicx, epsfig}

\newcommand{\be}{\begin{equation}}
\newcommand{\ee}{\end{equation}}
\newcommand{\bea}{\begin{eqnarray}}
\newcommand{\eea}{\end{eqnarray}}

\def\etal{{\it et al.}}

\begin{document}

\title{Is the low-$\ell$ microwave background cosmic?}

\author{Dominik J. Schwarz$^1$}
\author{Glenn D. Starkman$^{1,2}$}
\author{Dragan Huterer$^2$}
\author{Craig J. Copi$^2$}
\affiliation{$^1$ Department of Physics, CERN, Theory Division, 
             1211 Geneva 23, Switzerland}
\affiliation{$^2$ Department of Physics, Case Western Reserve University, 
             Cleveland, OH~~44106-7079}

\begin{abstract}
The large-angle (low-$\ell$) correlations of the Cosmic Microwave Background 
exhibit several statistically significant anomalies compared to the standard 
inflationary cosmology. We show that the quadrupole plane and the three 
octopole planes are far more aligned than previously thought ($99.9$\% C.L.).
Three of these planes are orthogonal to the ecliptic at $99.1$\% C.L., and 
the normals to these planes are aligned at $99.6$\% C.L.~with the direction 
of the cosmological dipole and with the equinoxes. The remaining octopole 
plane is orthogonal to the supergalactic plane at $99.6$\% C.L. 
\end{abstract}

\pacs{98.80.-k}

\maketitle

Much effort is currently being devoted to examining the cosmic microwave 
background (CMB) temperature anisotropies measured with the Wilkinson 
Microwave Anisotropy Probe (WMAP)~\cite{wmap_results,wmap_foreground,
wmap_angps,wmap_low} and other CMB experiments~\cite{otherCMB}. 
While the data is regarded as a dramatic confirmation of 
standard inflationary cosmology, anomalies exist. In particular the 
correlations at large angular separations, or low $\ell$, exhibit several 
peculiarities.

Most prominent among the ``low-$\ell$ anomalies'' is the near vanishing of 
the two-point angular correlation function $C(\theta)$ at angular separations 
greater than about 60 degrees. This was first measured using the Cosmic 
Background Explorer's Differential Microwave Radiometer (COBE-DMR) \cite{DMR4} 
and recently confirmed 
by observations with WMAP \cite{wmap_low}.
This anomalous lack of large-angle correlation is connected to the low value 
of the quadrupole contribution, $C_2$, in a spherical harmonic expansion of 
the CMB sky -- 
\be
\Delta T \left(\theta,\phi\right) \equiv 
\sum_{\ell=1}^{\infty} T_\ell \equiv 
\sum_{\ell=1}^{\infty}\sum_{m=-\ell}^{\ell} a_{\ell m} 
Y_{\ell m}\left(\theta,\phi\right)
\ee
(with $(2\ell+1)C_\ell=\sum_m \vert a_{\ell m}\vert^2$) -- although the 
smallness of $C_2$ does not fully account for the shape of $C(\theta)$. The 
significance of the large-angle behaviour of $C(\theta)$, especially in light 
of the large cosmic variance in $C_2$, is a matter of some controversy. 
The comparisons are, moreover, confused  by the fact that where one
author may calculate only the probability of the low value of $C_2$,
others, such as the WMAP team, calculate the probability of $C(\theta)$ 
being as low as it is over some range of angles. The issue of what prior 
probabilities and estimators \cite{efstathiou,Eriksenc,slosar_a} to use 
further complicates the statistical situation.

While the overall absence of large-scale power has attracted the most 
attention, several other large-angle anomalies have been pointed out.
De Oliveira-Costa \etal\ \cite{angelica} have shown that the octopole 
is unusually planar, meaning that the hot and cold spots of the octopolar
anisotropies lie nearly in a plane.
The same authors found that the axes ${\bf n_2}$ and ${\bf n_3}$ 
about which the ``angular momentum" dispersion 
$\sum_m m^2 \vert a_{\ell m}\vert^2$ of the quadrupole and octopole are 
maximized are unusually aligned, $\vert{\bf n_2} \cdot {\bf n_3}\vert = 0.9838$.
Eriksen \etal\ \cite{NorthSouth} found that the deficit in 
large-scale power is due to a systematic deficit in power between $\ell=2$ 
and $40$ in the north ecliptic hemisphere compared to the south 
ecliptic hemisphere.
Some of us \cite{vectors} have shown that the $\ell=4$ to $8$ multipoles 
exhibit an odd, but very unlikely ($\sim 1$\% probability), correlation with 
$\ell=2$ and $\ell=3$.
These low-$\ell$ anomalies (and others \cite{OtherAnomalies})
have all been pointed out before, but no simple connection has been made 
between them. Here we remedy that situation. 

By far the largest signal in the CMB anisotropy is the dipole, recently 
measured by WMAP \cite{wmap_results} to be $(3.346\pm 0.017)$mK in the 
direction 
$(l=263^\circ\!\!.85\pm0^\circ\!\!.1, b=48^\circ\!\!.25\pm0^\circ\!\!.04)$ 
in galactic coordinates. This is  
caused by the motion of the Sun with respect to the rest frame 
defined by the CMB. As shown by Peebles and Wilkinson \cite{dipole}, the dipole
induced by a velocity $v$ is $\bar{T} (v/c) \cos\theta$, where $\theta$ is 
measured from the direction of motion. Given $\bar{T}=(2.725\pm0.002)$K 
\cite{Mather}, one infers that $v\simeq 370 {\rm km}/{\rm s}$.

The solar motion also implies \cite{dipole,kineticquad_ab,kineticquad_c} 
the presence of a kinematically induced Doppler quadrupole (DQ). To second 
order in $\beta \equiv v/c \simeq 10^{-3}$, 
an observer moving with respect to the CMB rest-frame 
sees the usual monopole term with a black-body spectrum,  
a dipolar term $\propto \beta \cos\theta$ with a dipole spectrum 
and a quadrupolar term $\propto \beta^2 (3\cos^2\theta - 1)$ 
with a quadrupole spectrum. 
Higher multipoles are induced only at 
higher order in $\beta$ and so can be neglected. 
%
To first 
approximation the quadrupole spectrum differs very little from the dipole 
spectrum across the frequency range probed by WMAP. The DQ is itself a 
small contribution to the quadrupole. It has a total band-power 
of only $3.6\mu{\rm K}^2$ compared to $(154\pm70)\mu{\rm K}^2$ from the 
cut-sky WMAP analysis \cite{wmap_angps}, $195.1\mu{\rm K}^2$ extracted 
\cite{angelica} from the WMAP Internal Linear Combination (ILC) full-sky 
map \cite{wmap_products}, or $201.6\mu{\rm K}^2$ from the Tegmark 
\etal\ full-sky map \cite{tegmark_cleaned} (henceforward the Tegmark map). 
Therefore, it is a good approximation to treat the 
Doppler-quadrupole as having a dipole spectrum plus a small spectral 
distortion which we shall ignore. We can then readily subtract the DQ from 
the ILC or Tegmark map. (The ILC and Tegmark maps differ in the amount of 
spatial filtering used to produce them.) 


Meanwhile, some of us showed \cite{vectors} that the $\ell$-th multipole, 
$T_\ell$, can instead be written uniquely in terms of a scalar $A^{(\ell)}$
which depends only on the total power in this multipole (i.e. on $C_\ell$)
and $\ell$ unit vectors $\{ {\bf \hat v}^{(\ell,i)}\vert i=1,...,\ell\}$.
These ``multipole vectors'' are entirely independent of $C_\ell$,
and instead encode all the information  about the phase relationships
of the $a_{\ell m}$.  Heuristically,
\be
T_{\ell} \approx 
A^{(\ell)} \prod_{i=1}^{\ell}({\bf \hat v}^{(\ell,i)}\cdot {\hat e}) \, ,
\ee
where ${\bf \hat v}^{(\ell,i)}$ is the $i^{\rm th}$ multipole vector of the 
$\ell^{\rm th}$ multipole. (In fact the right hand side contains terms with 
``angular momentum" $\ell-2$, $\ell-4$, ... These are subtracted by taking 
the appropriate traceless symmetric combination, as described in 
\cite{vectors}.) Note that the signs of all the vectors can be 
absorbed into the sign of $A^{(\ell)}$. For convenience we take 
the vectors to point in the north galactic hemisphere. The multipole vectors 
for $\ell=2$ and $3$ for the DQ-corrected Tegmark map are [in galactic $(l,b)$]
\bea
{\bf \hat v}^{(2,1)} &=& (11^\circ\!\!.26,16^\circ\!\!.64),\nonumber \\
{\bf \hat v}^{(2,2)} &=& (118^\circ\!\!.87,25^\circ\!\!.13),\nonumber \\
{\bf \hat v}^{(3,1)} &=& (22^\circ\!\!.63,9^\circ\!\!.18),\\
{\bf \hat v}^{(3,2)} &=& (86^\circ\!\!.94,39^\circ\!\!.30),\nonumber \\
{\bf \hat v}^{(3,3)} &=& (-44^\circ\!\!.92,8^\circ\!\!.20).\nonumber 
\eea
(A similar analysis has been done for the ILC map. Results are quoted where 
instructive.)

\begin{figure}[t]
\label{fig:normals}
\includegraphics[width=\linewidth]{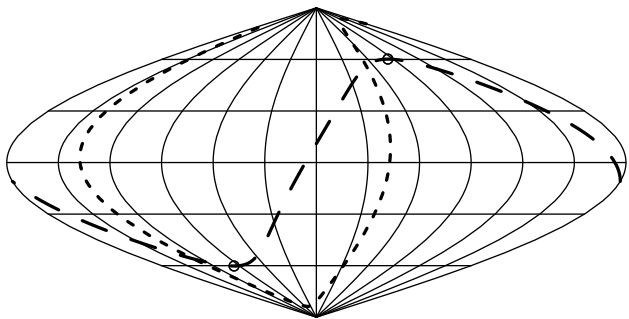}\\
\caption{The normal vectors for $\ell=2$ (
dark grey), and $\ell=3$ (
black), as well 
as $\pm$ the dipole direction (light grey), and the equinoxes (open circles)
plotted in sinusoidal projection. The Galactic center is the coordinate origin. 
Galactic longitude is positive on the left. Grid lines are separated by 
$30^\circ$. The long-dashed line is the ecliptic; the short-dashed line is 
the supergalactic plane. The clustering of the $\ell=3$ normal vectors
is indicative of the ``planarity of the octopole'' \cite{angelica}. (The 
clustering is clearer to the eye in the south hemisphere because of the 
projection.) Their closeness to the $\ell=2$ normal is indicative of the 
quadrupole-octopole correlation. Note how three normals are close to the 
dipole, the ecliptic and the equinoxes, while the fourth is on the 
supergalactic plane.
}
\end{figure}

As described in \cite{vectors} there are several ways to compare the 
multipole vectors, however most striking is to compute for each $\ell$ 
the $\ell(\ell-1)/2$ independent cross-products. These are the oriented areas
${\mathbf w}^{(\ell,i,j)} \equiv \pm\left({\bf \hat v}^{(\ell,i)} \times 
{\bf \hat v}^{(\ell,j)}\right)$. 
The overall signs of the area vectors are again unphysical (we take them 
to point in the north galactic hemisphere), however their magnitudes are not.
The area vectors for $\ell=2,3$ for the Tegmark map (cf.~Fig.~1) are 
\bea
{\mathbf w}^{(2,1,2)} &=& 0.9900(-105^\circ\!\!.73,56^\circ\!\!.62), \nonumber \\
{\mathbf w}^{(3,1,2)} &=& 0.9017(-78^\circ\!\!.38,49^\circ\!\!.76), \\
{\mathbf w}^{(3,2,3)} &=& 0.9072(-141^\circ\!\!.61,38^\circ\!\!.96), \nonumber \\
{\mathbf w}^{(3,3,1)} &=& 0.9184(173^\circ\!\!.77,79^\circ\!\!.54).\nonumber  
\eea
The directions of ${\mathbf w}^{(2,1,2)}$ and of de Oliveira-Costa \etal's ${\bf n}_2$ 
are mathematically equivalent. The small difference
is due to the removal of the DQ here in ${\mathbf w}^{(2,1,2)}$.

Finally, the magnitudes of the dot products between ${\mathbf w}^{(2,1,2)}$ and 
${\mathbf w}^{(3,i,j)}$ ordered from largest to smallest are:
\bea
A_1 \equiv \vert{\mathbf w}^{(2,1,2)} \cdot {\mathbf w}^{(3,1,2)}\vert &= 0.8509, 
\nonumber \\ 
A_2 \equiv \vert{\mathbf w}^{(2,1,2)} \cdot {\mathbf w}^{(3,2,3)}\vert &= 0.7829, \\
A_3 \equiv \vert{\mathbf w}^{(2,1,2)} \cdot {\mathbf w}^{(3,3,1)}\vert &= 0.7616. 
\nonumber 
\eea
Using instead the normal vectors ${\bf \hat n}^{(\ell,i,j)} \equiv 
{\mathbf w}^{(\ell,i,j)}/\vert{\mathbf w}^{(\ell,i,j)}\vert$, the dot products are:
\be D_1 = 0.9531, \quad D_2 = 0.8719, \quad D_3 = 0.8377. \ee
One can see from the large values of all three of the $D_i$ that the quadrupole
and octopole are aligned, and that the octopole is unusually planar.

The $A_i$ retain information about both the magnitudes and orientations of 
the ${\bf w}^{(\ell,i,j)}$. We have compared their values against $10^5$ 
Monte Carlos (MCs) of Gaussian random statistically isotropic skies with 
pixel noise (as in \cite{vectors}). The probability that the largest of 
these dot products is at least $A_1$, the $2^{\rm nd}$ largest at least $A_2$, 
and the $3^{\rm rd}$ largest at least $A_3$ is $0.021$\%. It is $0.11$\% 
without the DQ-correction, supporting that this is not just 
a statistical accident. (The ILC map yields $0.025$\% and $0.12$\%.) 

Ordering dot products does
not induce a well-defined ordering relation for the MC maps \cite{KW}. A 
robust and more conservative statistic is the sum of the dot products, 
$S = \sum_i A_i$. We find that only $0.128\%$ of the MC maps have a larger 
$S$ than the DQ-corrected Tegmark map. 
Thus {\em the quadrupole-octopole correlation is excluded from being a chance 
occurrence in a gaussian random statistically isotropic sky at 
$>99.87$\% C.L.} 
This result is statistically independent of (though perhaps not physically 
unrelated to) the lack of power at large angular scales since all the 
information about the power is contained in the $A^{(\ell)}$, and not in the 
multipole vectors. 

\begin{figure}[t]
\label{fig:TegmarkT23}
\includegraphics[width=\linewidth]{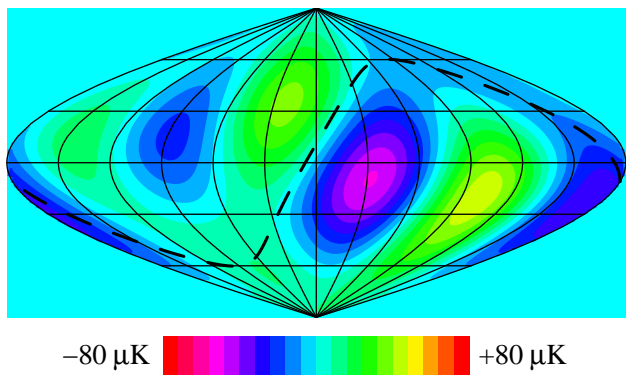}
\caption{Cosmic quadrupole plus octopole, $T^{\rm DS}_{2+3}$ (from Tegmark 
map). Coordinates are as in Fig.~1,  
background color is $0\mu$K. 
The ecliptic (dashed line) avoids all extrema. 
}
\end{figure}

So far we have looked only at the correlation between the CMB quadrupole and
octopole. Motivated by the results of Eriksen \etal\ \cite{NorthSouth}, we 
next ask whether the quadrupole and octopole correlate with the ecliptic 
or the galaxy. We notice (see Fig.~1) that three of the four 
${\bf \hat n}^{(\ell,i,j)}$ seem to lie near the ecliptic plane. Their dot 
products with the north ecliptic pole are (for the Tegmark map) --  in order 
of size: $0.0271$, $0.0450$, $0.1786$ and $0.5233$. This means that three 
of the four planes defined by the quadrupole and octopole are nearly orthogonal
to the ecliptic. The probability that a MC map has the same four dot products 
smaller than these is $0.104$\% (for the ILC map $0.193$\%). The sum of the 
four dot products, $S$, of MC maps exceeds that of the DQ-corrected Tegmark 
map only in $0.925$\%, so {\em a chance alignment of the normals with the 
ecliptic plane is excluded at $> 99$\% C.L.} 
A similar comparison to the galactic plane is unremarkable -- the normal 
closest to the galctic pole, $n^{(3,3,1)}$  (the normal not near the ecliptic),
could be even closer to the pole at $6.6$\% probablility; however, $n^{(3,3,1)}$
is only $0^\circ\!\!.07$ off the supergalactic plane (see Fig.~1). 
Consequently, the likelihood of the dot product of one $n^{(\ell,i,j)}$ with 
the supergalactic pole being as small as observed is just $0.449$\%.  

The three normals that are near the ecliptic also lie very near the 
axis of the dipole. The likelihood of this alignment with the dipole 
is $0.041$\% ($0.098$\%) for the Tegmark (ILC) map and 
$0.405$\% with respect to the $S$ statistics.  
Since the dipole axis lies so close to the equinoxes, this may also be 
viewed as an alignment with the equinoxes. The angular difference 
in ecliptic longitude of the normals from the equinox is $<1^{\rm d}$, 
$13^{\rm d}$, $13^{\rm d}$, and $44^{\rm d}$. The probability of these four 
headless vectors having ecliptic longitudes this close to the equinoxes is 
$0.039$\% ($0.005$\%) for the Tegmark (ILC) map, $0.368$\% for the 
$S$ statistics. This is statistically independent from their 
near orthogonality to the ecliptic poles.

Two different statistical test (ordered dot products and sum of dot products)
show evidence for low-$\ell$ anomalies in the Tegmark map (and consistent 
results for the ILC map). These anomalies could be due to a wrong treatment 
of foregrounds, although these full-sky maps do not refer to a foreground 
model. However, 
 they employ a linear combination of
 frequencies to produce a minimum variance map. They are 
 susceptible to bias because chance alignments between CMB and
 Galactic signals tend to cancel in order to produce minimum variance.
Error estimates for the ILC and Tegmark maps have not been published, 
nevertheless we try to estimate the errors on the correlations induced by 
errors in the full-sky maps. As we add (ad hoc) Gaussian noise of $10 \mu$K to 
$a_{20}$ (which has a value of $9.3 \mu$K in the Tegmark map), the evidence (from ordered dot 
products) for the quadrupole-octopole alignment is slightly reduced, 
from $99.97$\% C.L.\ 
to $99.57$\% C.L. All other correlations remain at $>99$\% C.L. Adding the 
same error to all $a_{\ell m}$ with $\ell = 2$ and $3$ preserves all 
correlations at $> 95$\% C.L.

In 
Fig.~2, we plot a combined Doppler-subtracted quadrupole 
plus octopole map, $T^{DS}_{2+3}$. 
The dashed line is the ecliptic. Strikingly, the 
ecliptic threads its way along the node line separating one of the hot spots
from one of the cold spots, tracking the node over a 
third of the sky (and avoiding the extrema over the rest).  
The alignment is even better in the ILC map (not shown) than in the 
Tegmark map, which 
is not surprising given the spatial filter used by Tegmark \etal\
Replotting in other projections (not shown) does not noticeably 
alter the apparent correlation. A second look at Fig.~2 
reveals a north-south ecliptic asymmetry -- the three extrema in the 
north are visibly weaker than those in the south. One is cautioned that, 
given the observed planarity of the quadrupole-plus-octopole, one expects some 
hemispheric asymmetry because of the parity of the octopole. This may explain 
the asymmetry found in \cite{NorthSouth}.

Interestingly, the correlation with the ecliptic is stronger in the combined 
quadrupole-octopole map, than in either 
separately,
and the north-south ecliptic asymmetry is visible in neither alone. The 
correlation (but not the asymmetry) appears to be weakened if higher 
multipoles are added in, however a more thorough analysis is merited.
(For example, in $T_4$ the extrema seem to define
both the ecliptic and the plane of the quadrupole-octopole, and the dipole 
direction lies near the center of a cold spot.)
These trends are exhibited in Fig.~3. 
\begin{figure}[t]
\label{fig:T2T3T234}
\includegraphics[width=0.32\linewidth]{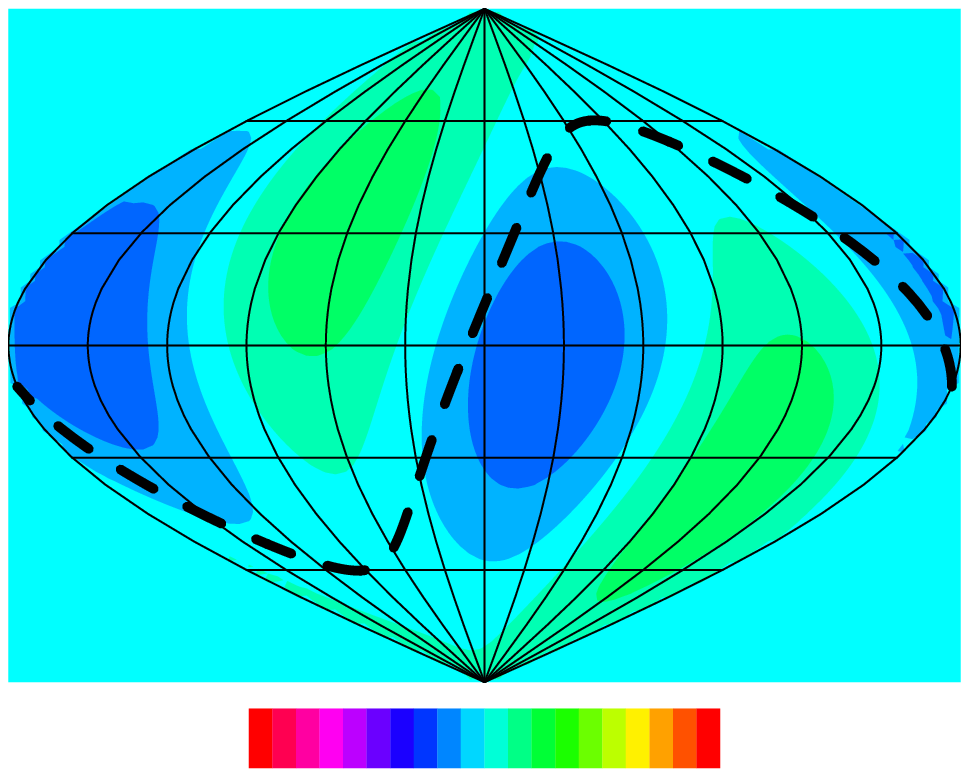}
\includegraphics[width=0.32\linewidth]{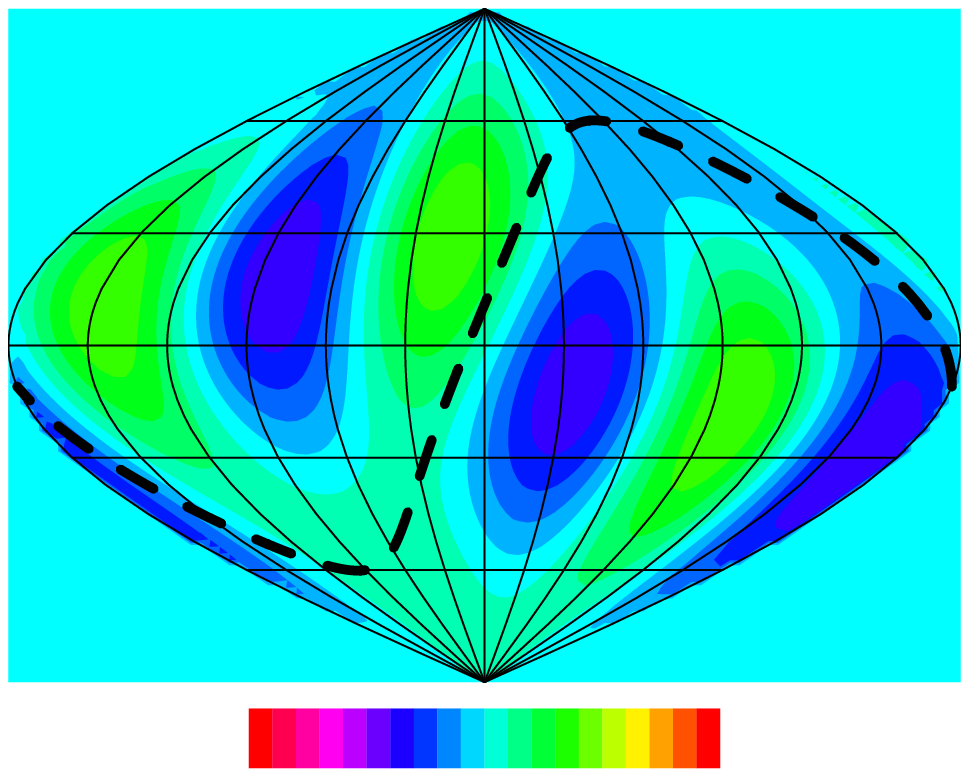}
\includegraphics[width=0.32\linewidth]{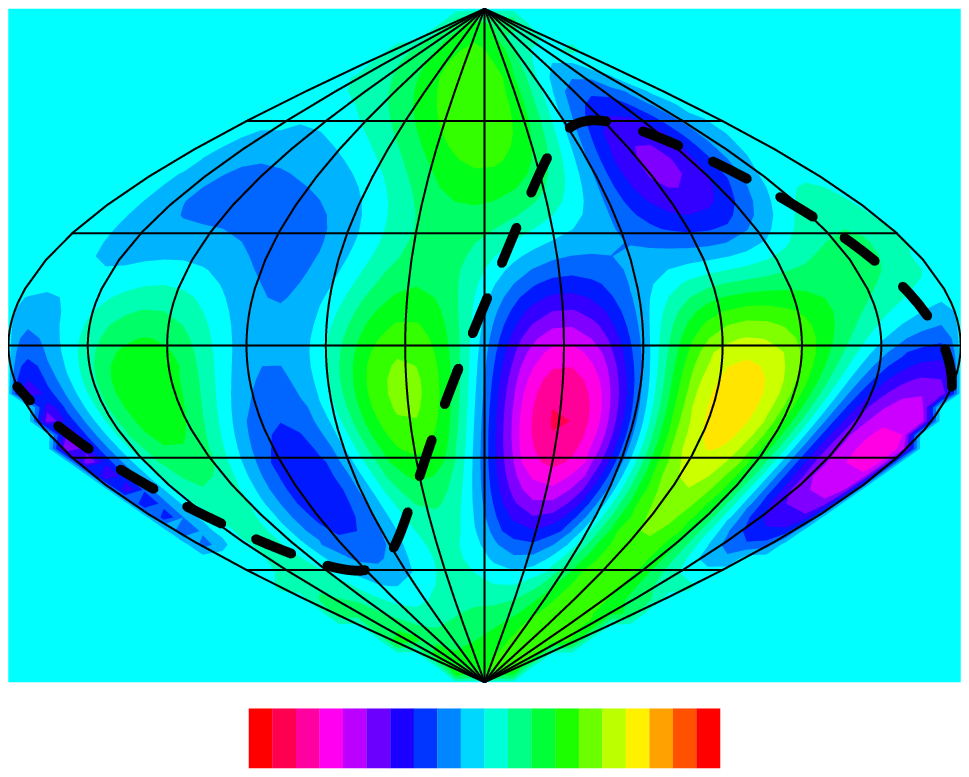}\\
\caption{Left to right: Cosmic (Doppler-subtracted) quadrupole, octopole, and 
$T^{\rm DS}_{2+3}+T_4$ from the Tegmark map. The 
coordinate system and color scale are as in Fig.~2. 
The ecliptic (dashed line) is noticeably less correlated with 
these maps than with the quadrupole-octopole map of Fig.~2. 
}
\end{figure}

We have shown that the planes defined by the octopole are nearly aligned 
with the plane of the Doppler-subtracted quadrupole, that three of these 
planes are orthogonal to the ecliptic plane, with normals aligned with 
the dipole (or the equinoxes), while the fourth plane is perpendicular to the 
supergalactic plane.  Each of these correlations is unlikely at 
$\geq 99$\% C.L., and at least two of them are statistically independent.
We have also seen that the ecliptic threads between a hot and a cold spot 
of the combined Doppler-subtracted-quadrupole and octopole map -- following 
a node line across about $1/3$ of the sky, and separating the three strong 
extrema from the three weak extrema of the map. 

{\it We find it hard to believe that these correlations are just statistical 
fluctuations around standard inflationary cosmology's prediction of 
statistically isotropic Gaussian random $a_{\ell m}$s.}  That the 
quadrupole-octopole correlation just happened to increase by $\sim5$ when the 
quadrupole was Doppler-corrected seems particularly unlikely.  
The correlation of the normals with the ecliptic poles 
suggest an unknown source or sink of CMB radiation or an unrecognized 
systematic.  If it is a physical source or sink in the inner-solar system 
it would cause an annual modulation in the time-ordered data.  An outer 
solar-system origin (beyond $100$ A.U.) might avoid such a signal. It seems
likely that a local source (sink) would also show up in polarization maps,
especially there would be no reason for B-modes being significantly 
suppressed with respect to E-modes. 
Physical correlation of the CMB with the equinoxes is difficult to imagine, 
since the WMAP satellite has no knowledge of the inclination of the Earth's 
spin axis.  Alternatively, this normals-ecliptic correlation could be a 
consequence of the closeness of the dipole to the ecliptic plane and the 
correlation of the normals to the dipole.
But since the dipole is itself believed to be due to local motion,
this would suggest non-cosmological contamination of $\ell=2$ and $3$,
as does the observed correlation with the supergalactic plane.

Although full-sky maps such as the ones we have used are not expected 
to be reliable at high l and do not agree with each other with respect to 
the galaxy, for the existing maps (ILC map, Tegmark map and Eriksen 
\etal\ map [8]) the ecliptic plane does not show up in the difference maps 
and their power spectra are consistent with each other at low-l \cite{Eriksenc}.

If indeed the $\ell=2$ and $3$ CMB fluctuations are 
inconsistent with the predictions of standard cosmology,
then one must reconsider all CMB  results within the standard 
paradigm which rely on low $\ell$'s, including: 
the high temperature-polarization correlation $C_\ell^{TE}$ 
measured by WMAP \cite{wmap_results} at very low $\ell$ 
(and hence the inferred redshift of reionization);
the normalization of the primordial fluctuations (which 
relies on the extraction of the optical depth $\tau$ from low $\ell$ ); 
and the running $d n_s/d\log{k}$ of the spectral index  of scalar perturbations 
(which, as noted in \cite{VSA}, depends on the absence of low-$\ell$ TT power).

The authors thank H.K.\ Eriksen, S.M.\ Leach, A.\ Lue, U.\ Seljak, D.\ Spergel, 
J.\ Steinberger, I.\ Tkachev, T.\ Vachaspati and J.\ Weeks for useful 
suggestions and discussions.  The work of GDS, DH and CJC has been supported 
by the US DoE. GDS acknowledges Maplesoft for the use of Maple V. We have 
benefited from using the publicly available Healpix package~\cite{healpix}. 


\end{document}